\documentclass[aps,twocolumn,groupedaddress,showpacs]{revtex4}
\usepackage[pdftex]{graphicx}
\newcommand{\isum}%
{\mathop{\hbox{$\displaystyle\sum\kern-13.2pt\int\kern1.5pt$}}}
\renewcommand{\r}{{\bm r}}
\renewcommand{\k}{{\bm k}}
  
   \newcommand{\n}{{\bm n}}

  \newcommand{\mc}{\multicolumn}
  
\newcommand{\bt}{\begin{tabular}}
\newcommand{\et}{\end{tabular}}

\newcommand{\eref}[1] {(\ref{#1})}
\newcommand{\Eref}[1] {Eq.~(\ref{#1})}

\newcommand{\Fref}[1] {Figure \ref{#1}}

\newcommand{\np}{\newpage}

\newcommand{\br}{\begin{eqnarray*}}
\newcommand{\er}{\end{eqnarray*}}
\newcommand{\ba}{\begin{eqnarray}}
\newcommand{\ea}{\end{eqnarray}}
\newcommand{\be}{\begin{equation}}
\newcommand{\ee}{\end{equation}}

\newcommand{\vs}{\vspace*}
\newcommand{\bp}{\begin{minipage}}
\newcommand{\ep}{\end{minipage}}

\begin{document}
\bibliographystyle{apsrev}

\title {Strong-field ionization of He by elliptically polarized light
  in attoclock configuration }

\author{I. A. Ivanov }
\email{Igor.Ivanov@anu.edu.au}
\author{ A. S. Kheifets}
\email{A.Kheifets@anu.edu.au}

\affiliation{Research School of Physical Sciences,
The Australian National University,
Canberra ACT 0200, Australia}

\date{\today}
\begin{abstract}
  We perform time-dependent calculations of strong-field ionization of He by
  elliptically polarized light in configuration of recent attoclock measurements
  of Boge {\em et al} [PRL {\bf 111}, 103003 (2013)].  By solving a 3D
  time-dependent Schr\"odinger equation, we obtain the angular offset $\theta_m$
  of the maximum in the photoelectron momentum distribution in the polarization
  plane relative to the position predicted by the strong field
  approximation. This offset is used in attoclock measurements to extract the
  tunneling time. Our calculations clearly support the set of experimental
  angular offset values obtained with the use of non-adiabatic calibration of
  the {\em in situ} field intensity, and disagree with an alternative set
  calibrated adiabatically.  These findings are in contrast with the conclusions
  of Boge {\em et al} who found a qualitative agreement of their semiclassical
  calculations with the adiabatic set of experimental data. This controversy may
  complicate interpretation of the recent atto-clock measurements.

\end{abstract}

\maketitle

%\section{Introduction}

One of recent advances of attosecond science was experimental observation of the
time delay of photoemission after subjecting an atom to a short and intense
laser pulse. Theoretical interpretation of such measurements depends on the
Keldysh parameter $\gamma$ which draws the borderline between the truly quantum
multiphoton regime $\gamma>1$ and a semi-classical tunneling regime $\gamma<1$
\cite{Keldysh64}. The time delay measurements in the multi-photon regime by
attosecond streaking \cite{M.Schultze06252010} or two-photon sideband
interference \cite{PhysRevLett.106.143002,PhysRevA.85.053424} can be
conveniently interpreted by the Wigner time delay theory
\cite{deCarvalho200283}. Even though some quantitative differences remain
between measured and calculated time delays (see
e.g. \cite{PhysRevA.87.063404}), qualitatively, these measurements are now well
understood. At the same time, interpretation of the attosecond measurements in
the tunneling regime by attosecond angular streaking
\cite{Eckle2008,P.Eckle12052008} or high harmonics generation \cite{Shafir2012}
is less straightforward. Indeed, the timing of the tunneling process has been a
subject of numerous discussions and a long controversy (see
\cite{RevModPhys.66.217} for a comprehensive review).

The attosecond angular streaking technique, termed colloquially as a tunneling
clock or an atto-clock, uses the rotating electric-field vector of the
elliptically polarized pulse to deflect photo-ionized electrons in the angular
spatial direction. Then the instant of ionization is mapped to the final angle
of the momentum vector in the polarization plane, and a tunneling time is
calculated using a semiclassical propagation model.  By employing this
technique, \citet{P.Eckle12052008} placed an intensity-averaged upper limit of
12~as on tunneling time in strong field ionization of He with peak intensities
ranging from 2.3 to 3.5 units of $\rm 10^{14}~W/cm^2$.
In a subsequent paper by the same group \cite{Pfeiffer2012}, the attoclock was
used to obtain information on the electron tunneling geometry and to confirm
vanishing tunneling time. In addition, by comparing the angular streaking
results in Ar and He, multi-electron effects were clearly identified. Further
on, the influence of the ion potential on the departing electron was considered
and explained within a semiclassical model
\cite{PhysRevLett.109.083002,0953-4075-46-12-125601}.
In a recent development \cite{2013arXiv1306.6280W}, the attoclock technique was
transferred from a cold-target recoil-ion momentum spectrometer ({\sc coltrims})
to a velocity map imaging spectrometer ({\sc vmis}).  These refined attoclock
measurements revealed a real and not instantaneous tunneling time over a large
intensity regime \cite{2013arXiv1301.2766L}. Various competing theories of
tunneling ionization were assessed against these experimental data, and some of
them were found consistent with the data.

In the latest report \cite{PhysRevLett.111.103003}, the attoclock measurements
on He were used to assess the influence of non-adiabatic tunneling effects. In
the tunneling regime, the electron tunnels adiabatically, it experiences a
static field while tunneling and exits the tunnel with zero momentum
\cite{Keldysh64}.  By employing both the {\sc coltrims} and {\sc vmis}
techniques, the attoclock measurements of Ref.~\cite{PhysRevLett.111.103003}
were extended over a large range of intensities from 1 to 8 units of
$10^{14}~\rm W/cm^2$, corresponding to a variation of the Keldysh parameter
$\gamma$ from 0.7 to 2.5.  The upper end of the $\gamma$ interval clearly
trespasses on the multiphoton regime where the adiabatic hypothesis becomes
questionable and the electron exits the tunnel with a non-zero momentum. Because
this exit momentum is used as a tool for {\em in situ} calibration of the field
intensity in the attoclock experiments, adopting either of the adiabatic or
non-adiabatic tunneling hypothesis would affect strongly the intensity
calibration and the tunneling time results. In order to overcome this
uncertainty, \citet{PhysRevLett.111.103003} performed a measurement of the angle
of the photoelectron momentum at the detector defined by $ \theta_m=
\arctan(p_{x~\rm final}/p_{y~\rm final})$. Provided the electron is tunnel
ionized at the maximum of the electric field $E_x$ and is driven to the detector
by the laser pulse, its final momentum is aligned with the vector potential at
the moment of ionization $A_y$ and hence $\theta_m=0$. Non-zero values
$\theta_m\ne0$ can be attributed to the Coulomb field of the ionic core and/or a
finite tunneling time

\citet{PhysRevLett.111.103003} obtained two sets of the offset angles $\theta_m$
under the two tunneling scenarios. Then they attempted to reproduce their data
qualitatively with a TIPIS model (Tunnel Ionization in Parabolic coordinates
with Induced dipole and Stark shift). The version of the model based on the
non-adiabatic tunneling hypothesis predicted increasing of the offset angle with
increase of the field intensity.  Conversely, the adiabatic model showed
decrease of the offset with growing intensity, which was indeed the case
experimentally. On this qualitative basis, \citet{PhysRevLett.111.103003}
concluded that their experiments conformed to the adiabatic tunneling
scenario. Quantitative difference of the adiabatic experimental data and theory
was attributed to a finite tunneling time. Comparable difference between the
non-adiabatic TIPIS theory and experiment can also be attributed to the same
finite tunneling time effect \cite{Cirelli2013}.

In the present work, we perform accurate numerical calculations of the angular
offset $ \theta_m$ by solving a 3D time-dependent Schr\"odinger equation
(TDSE). Our theoretical model is fully {\em ab initio}, it uses no adjustable
parameters and does not require any specific tunneling hypothesis.  Results of
our calculations support the set of experimental data calibrated under the
non-adiabatic hypothesis. If this agreement is not accidental, it may indicate
the influence of non-adiabatic effects predicted by the analytical theory
\cite{PhysRevA.64.013409}. It may also raise a question of validity of the TIPIS
model and, more broadly, the interpretation of the tunneling time measurements
reported in \cite{2013arXiv1301.2766L}.
%%
%% This set of attoclock measurements was calibrated adiabatically. It showed a
%% large angular offsets $\theta_m$ of the photoelectron momentum distribution
%% which were converted to noticeable time delay values by subtracting the
%% theoretical predictions of the TIPIS model. The tunneling time was
%% proclaimed real as opposite to the imaginary time corresponding to the decay of
%% the wave function under the barrier \cite{Lein2012}.  The two theoretical
%% predictions were found compatible with these measurements: the Larmor time
%% \cite{PhysRevB.27.6178} and a probabilistic approach based on the Feynman path
%% integrals \cite{PhysRevLett.65.2321,PhysRevLett.93.170401}. 
%%
Indeed, our numerical results, the TIPIS model predictions and the experimental
data of \citet{PhysRevLett.111.103003} are mutually contradictory. The adiabatic
tunneling scenario leads to the experimental data calibration which contradicts
to the present calculation. The non-adiabatic scenario leads to the TIPIS model
prediction which is qualitatively incompatible with the experiment. One of the
components of this triad, formed by the two theories and the experiment, is
likely to be at fault.

Because of this important implication, we made every effort possible to verify
our theoretical model and to validate our numerical computations. We tested the
gauge invariance, partial wave and radial box convergence and the carrier
envelope phase (CEP) as well as the pulse length effects. All these tests were
performed successfully.

%\section{Theoretical model and numerical calculations}

\begin{figure}[h]
%\hs{-0.07\columnwidth}
%\epsfxsize=0.55\columnwidth
%\epsffile{e.eps}
%\epsffile{fig1a.eps}

%\hs{-0.08\columnwidth}
%\epsfxsize=0.55\columnwidth
%\epsffile{fig1b.eps}
%\epsffile{a.eps}
\includegraphics[width=\columnwidth]{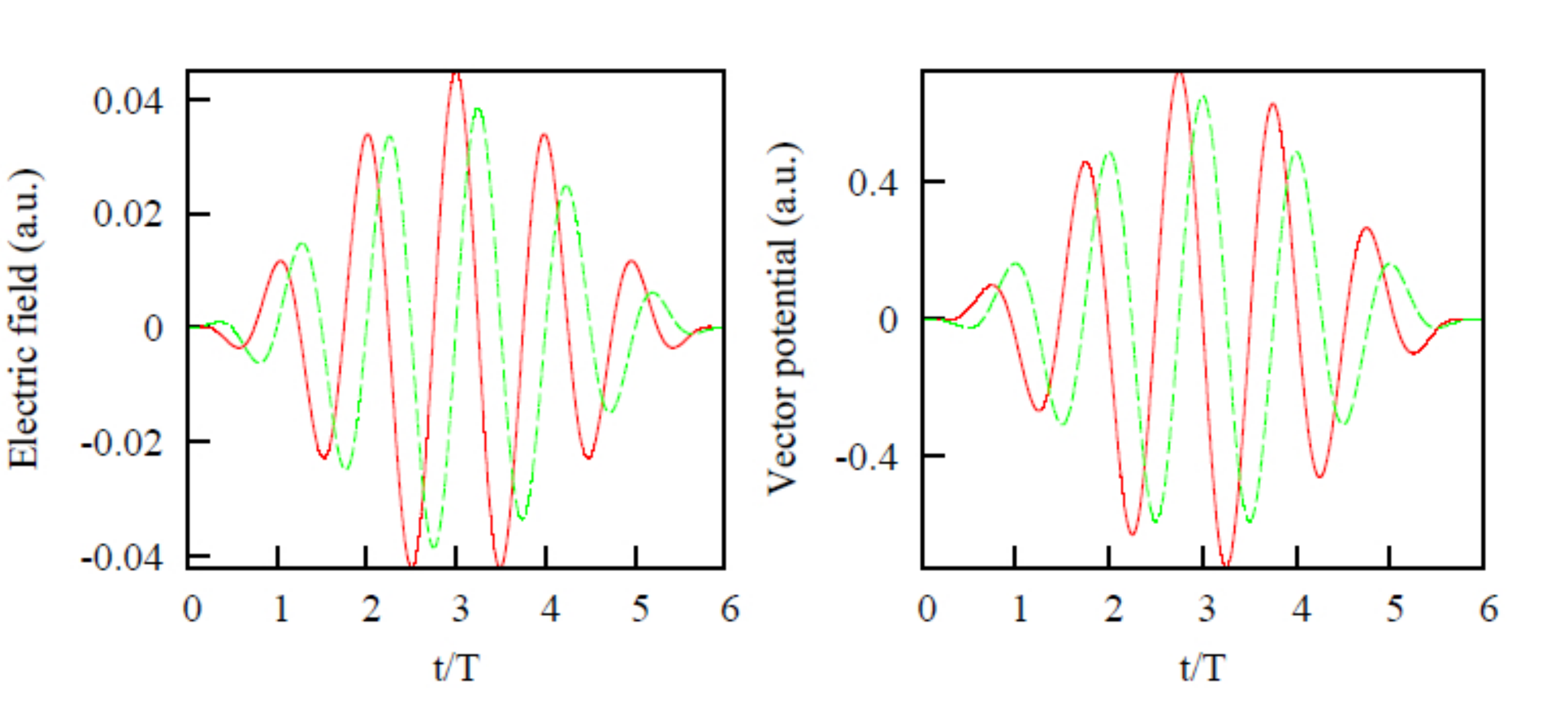}

\caption{(Color online) The electric field (left) and the vector potential
(right) of the laser pulse with $\phi=0$. Solid (red) line: $x$-components,
dashed (green): $y$-components.
\label{fig1}}
\end{figure}

We solve the TDSE for a helium atom described in a single active electron
approximation:
\begin{equation}
\label{TDSE}
i {\partial \Psi(\r) / \partial t}=
\left[\hat H_{\rm atom} + \hat H_{\rm int}(t)\right]
\Psi(\r) \ ,
\end{equation}
where $\hat H_{\rm atom}$ is the Hamiltonian of the field-free atom with
effective one-electron potentials \cite{Sarsa2004163,PhysRev.184.1}. Two
different model potentials were employed and produced indistinguishable results,
which assured the accuracy of the calculation.  The Hamiltonian
$\hat H_{\rm int}(t)$ describes the interaction with the EM field. For this
operator we can use both the length and velocity gauges:
\be
\label{gauge}
\hat H_{\rm int}(t) =
\left\{
\begin{array}{ccc}
{\bm E}(t) \cdot   \hat{\bm r} \\
 {\bm A}(t)\cdot \hat{\bm p} & , \ \  
{\bm A(t)}=-\int_{-T_1/2}^{t}{\bm E(\tau)}\ d\tau\\
\end{array}
\right.
\ee
The field is elliptically polarized in the $xy$ plane 
with the components:
\be
E_x= {{\cal E}f(t) \over \sqrt{1+\epsilon^2}} 
 \cos(\omega t +\phi) \  , \ 
E_y= {\epsilon {\cal E} f(t) \over \sqrt{1+\epsilon^2}} 
 \sin(\omega t+\phi) 
\label{ef}
\ee
Here the ellipticity parameter $\epsilon=0.87$ and the carrier frequency
$\omega=1.69$~eV (corresponding the wavelength $\lambda=735$ nm) are the same as
in the experimental work \cite{2013arXiv1301.2766L}. The pulse envelope was
chosen to be $f(t)= \sin^2\pi t/ T_1$, where $T_1=6T$ was the total pulse
duration ($T=2\pi/\omega$ is an optical period corresponding to the carrier
frequency), and $\phi$ the CEP.  The bulk of calculations was performed with
$\phi=0$ with a well-defined maximum of the vector potential relative to which
the angular offset is measured.  The electric field $\bm E$ and the vector
potential $\bm A$ of this pulse are shown in \Fref{fig1}. Some calculations at
few selected field intensities were performed with varying $\phi$. We also
performed a separate set of calculations at varying field intensity for a
shorter pulse with $T_1=3T$.

We seek a solution of \Eref{TDSE} in the form of a partial
wave expansion
\be
\Psi({\bm r},t)=
\sum\limits_{l=0}^{L_{\max}} \sum\limits_{\mu=-l}^{l}
f_{l\mu}(r,t) Y_{l\mu}(\theta,\phi)\ ,
\label{bas}
\ee
The radial part of the TDSE is discretized on a spatial grid in a
box.
To propagate the wave function \eref{bas} in time, we use
the matrix iteration method developed in
\cite{PhysRevA.60.3125} and further tested in calculations
of strong field ionization driven by linear \cite{klaus1,
dstrong} and circularly polarized \cite{PhysRevA.87.033407}
radiation.

By projecting the solution of the TDSE at the end of the laser pulse at
$t=T_1$ on the set of the ingoing scattering states:
\be
\psi^{(-)}_{\k}(\r)=
\sum\limits_{l\mu}
i^l e^{-i\delta_l}Y^*_{l\mu}(\n_{\k})Y_{l\mu}(\n_{\r})R_{kl}(r) \ ,
\label{partial}
\ee
(here $\n_{\k}=\k/k$, and $\n_{\r}=\r/r$ are unit vectors in
the direction of $\k$ and $\r$, respectively) we obtain
ionization amplitudes and the electron momentum
distribution:
\be
P({\k})=|\langle \psi^{(-)}_{\k}|\Psi(T_1)\rangle|^2
\label{proj}
\ee

%\section{Numerical tests}

For the  field parameters that we considered, the ionization
probabilities are extremely small (of the order of $10^{-10}$) which
required highly accurate computations.  The issue of convergence and
accuracy of the results was, therefore, critical for us
in the present work.
We found that convergence with respect to the number of partial waves retained
in \Eref{bas} is much faster in the velocity (V) gauge for the operator of the
atom-field interaction \eref{gauge}.  In the V-gauge, a convergence on the
acceptable level of accuracy was achieved for $L_{\max}=40$ (laser intensity of
$1.25\times 10^{14}$ W/cm$^2$ or less), $L_{\max}=50$ for the intensity of
$1.5\times 10^{14}$ W/cm$^2$, $L_{\max}=60$ for the intensities in the range
$1.75\times 10^{14}-2.25\times 10^{14}$ W/cm$^2$, and $L_{\max}=70$ for higher
intensities. In comparison, for the intensity of $1.25\times 10^{14}$ W/cm$^2$,
the L-gauge results begin to converge for $L_{\rm max}$ as large as 60. This
made use of the L-gauge for higher field intensities prohibitively expensive.
Results reported below, therefore, have been obtained using the V-gauge.
Typical calculation required several hundred hours of CPU time, which was only
possible by making our code run in parallel on a 1.2 petaflop supercomputer.  A
series of checks was performed to insure convergence both with respect to the
parameter $L_{\rm max}$, time integration stepsize $\Delta t$ and the box size
$R_{\rm max}$.  Some results of these checks are illustrated in Table I for the
field intensity of $1.25\times 10^{14}$ W/cm$^2$.  These checks allowed us to
estimate the error margin of our calculation as one degree.

\begin{table}[h]
\caption{\label{tab0}
Convergence with respect to 
the parameter $L_{\rm max}$ and the time integration stepsize $\Delta t$
for the field intensity of $1.25\times
10^{14}$ W/cm$^2$, $T_1=3T$ and $\phi=0$.}
\begin{ruledtabular}
\bt{cclc }
\mc{3}{l}{Computation parameters} & Ionization probability \\
Gauge & $L_{\rm max}$ & $\Delta t$, a.u & $10^{-10}$\\
\hline
\noalign{\smallskip}
V& 40& 0.01   &  1.0235\\
V& 50& 0.01   &  1.0115\\
V& 40& 0.0075 &  1.0234\\
L& 50& 0.01   &  0.807\\
L& 60& 0.01   &  0.959\\
\noalign{\smallskip}
\et
\end{ruledtabular}
\end{table}

\begin{figure}[h]
%%  \bt{cc}
%%  \vs{-0.5cm}
%%  \hs{-1.0cm} \epsfxsize=6cm\epsffile{Data/1e14/3D.eps} &
%%  \hs{-1.75cm}\epsfxsize=6cm\epsffile{Data/1.75e14/3D.eps} \\
%%  \vs{-0.55cm}
%%  \hs{-1.0cm} \epsfxsize=6cm\epsffile{Data/1.25e14/3D.eps} &
%%  \hs{-1.75cm}\epsfxsize=6cm\epsffile{Data/2e14/3D.eps} \\
%%  \vs{-0.55cm}
%%  \hs{-1.0cm} \epsfxsize=6cm\epsffile{Data/1.5e14/3D.eps} &
%%  \hs{-1.75cm}\epsfxsize=6cm\epsffile{Data/2.25e14/3D.eps} \\
%%  \et
%% \bs
%\epsfxsize=\columnwidth
%\epsffile{fig2.eps}
 \includegraphics[width=\columnwidth]{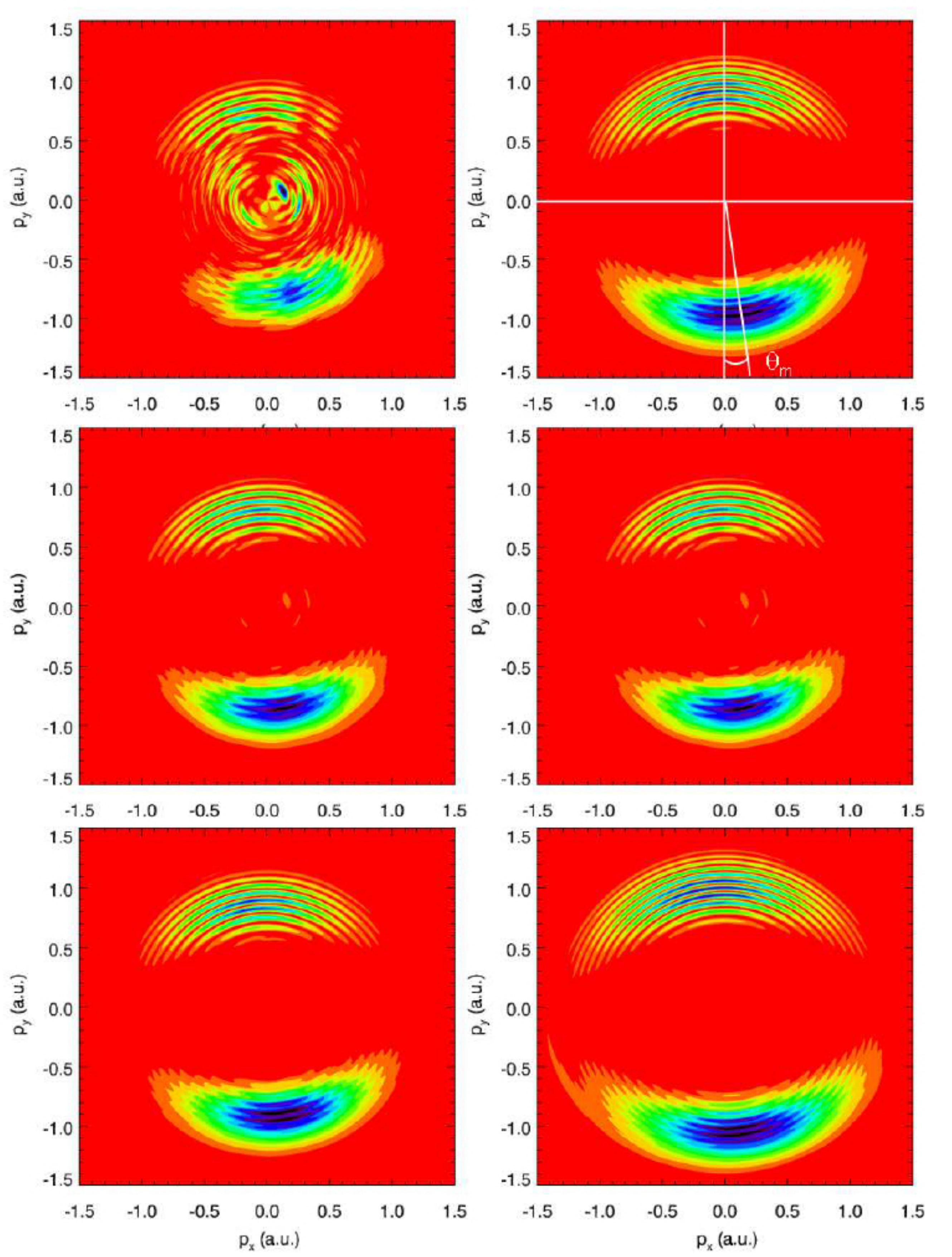}
%\epsffile{Fig01.eps}
\vs{-0.75cm}
\caption{(Color online) Photoelectron momentum distribution in the polarization
plane for the field intensities of 1, 1.25 and 1.5 units of $10^{14}$ W/cm$^2$
(left set of panels, from top to bottom) and 1.75, 2, and 2.25 units of
$10^{14}$ W/cm$^2$ (right set of panels, from top to bottom),  $T_1=6T$,
$\phi=0$.
The offset angle
$\theta_m$ relative to the vertical $-p_y$ direction is visualized on the top
right panel. 
\label{fig2}}
\end{figure}

\begin{figure}[h]
%\epsfxsize=6cm
%\epsffile{Data/PLOT.eps}\\
%\epsffile{fig3.eps}\\
 \includegraphics[width=6cm]{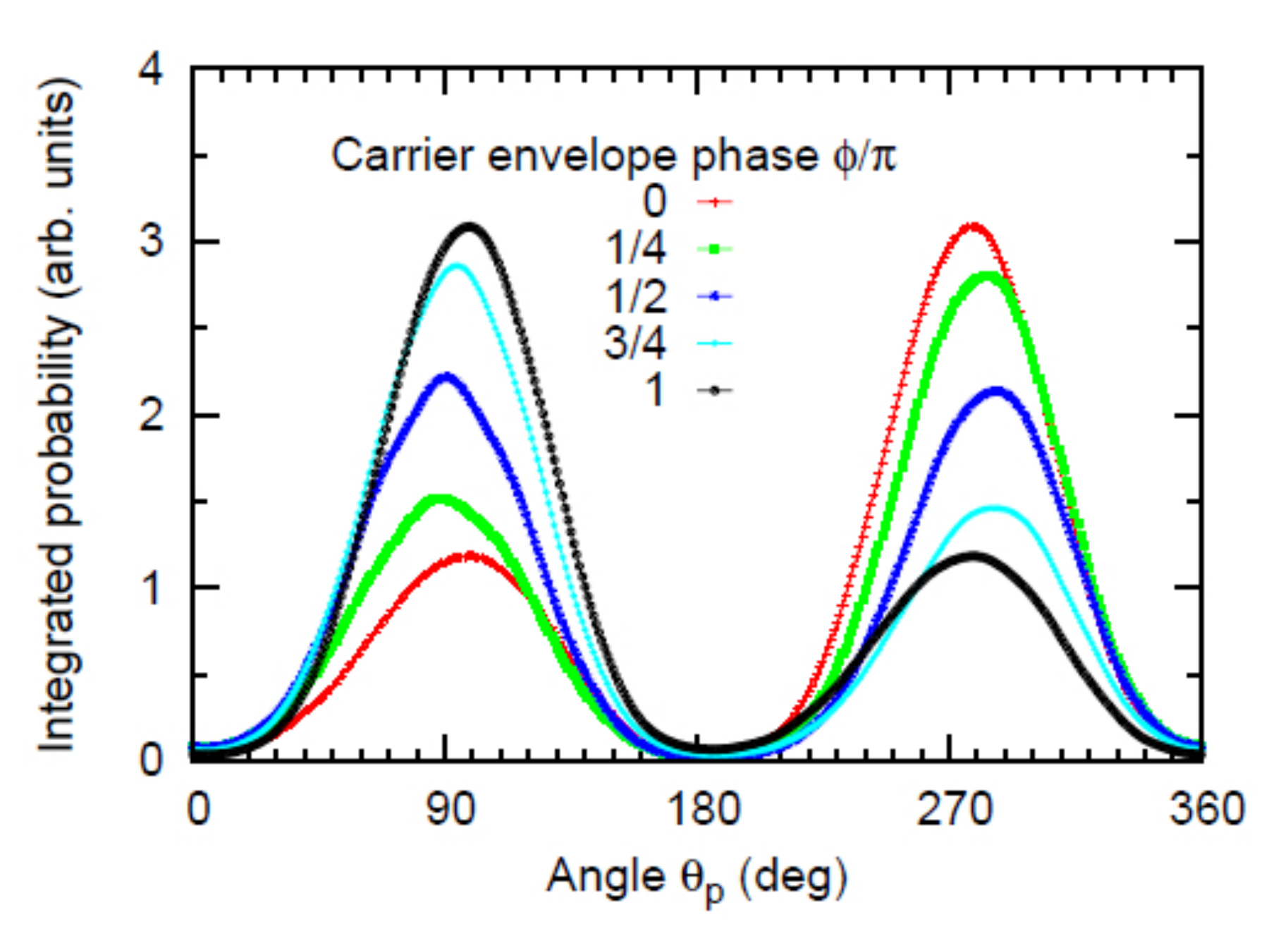}
\caption{(Color online) The radial $p$-integrated momentum distributions as the
functions of the angle $\theta_p$ for field intensities of 1, 1.25, 1.5, 1.75
and 2 units of  $10^{14}$ W/cm$^2$. 
The CEP $\phi=0$, pulse duration $T_1=6T$.
\label{fig3}}
\end{figure}

By using the projection operation \eref{proj}, we calculate the electron
momentum distribution in the polarization $xy$-plane. These distributions are
shown in \Fref{fig2} for the field intensities varying from $1\times10^{14}$ to
$2.25\times10^{14}$ W/cm$^2$.  Distributions were computed on a dense momenta
grid in the $p_xp_y$ plane using the polar coordinates $p$ and $\theta_p$. To
find the angular maximum $\theta_m$, we integrated the momentum distribution
over $p$ and analyzed resulting one-dimensional angular distribution. These
distributions for varying field intensities are shown in \Fref{fig3}.  A
similar procedure was followed in atto-clock experiments.

The well-known strong field approximation (SFA) \cite{Perelomov67} predicts that
the direction of the maximum of the momentum distribution in the polarization
plane should coincide with the direction of the vector potential $-\bm A(t_0)$
at the moment $t_0$ when the maximum field strength is attained.  For the pulse
with $T_1=6T$ and $\phi=0$, $t_0=3T$ which is the midpoint of the laser pulse.
The vector potential at this moment of time has zero $x$ and positive $y$
components (see the right panel of \Fref{fig1}). The SFA predicts, therefore the
zero offset angle $\theta_m=0$ from the vertical $-p_y$ direction. Our TDSE
calculations predict a noticeable offset $\theta_m$ relative to this direction
which is visualized on the top right panel of \Fref{fig2}.

\begin{figure}[h]
%\epsfxsize=6cm
%\epsffile{Data/1.5e14/PLOT.eps}
%\epsffile{fig4.eps}\\
 \includegraphics[width=6cm]{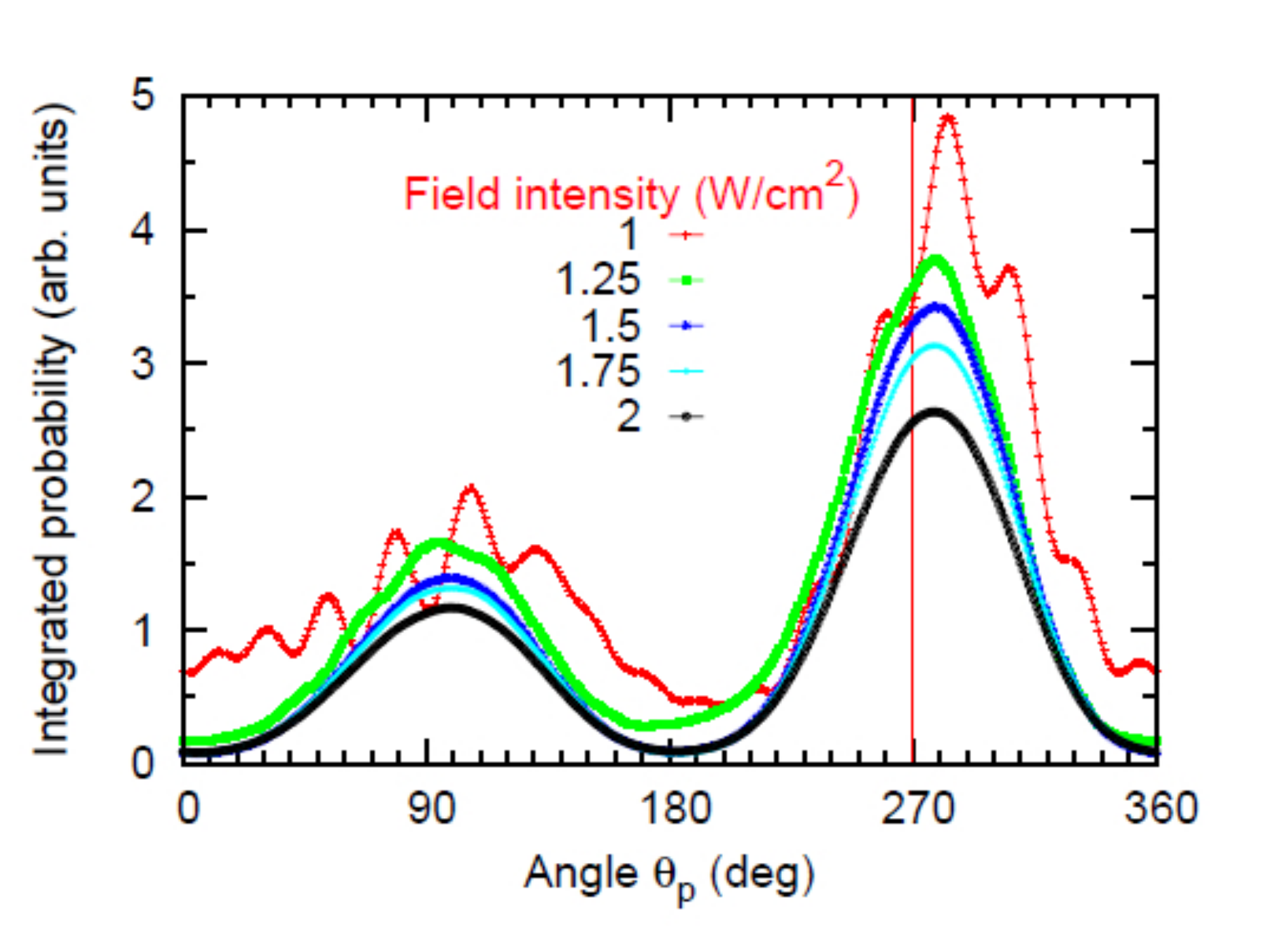}

\caption{(Color online) The radial $p$-integrated momentum distributions as the
functions of the angle $\theta_p$ for CEP values $\phi=0$, $\pi/4$,
$\pi/2$, $3\pi/4$ and $\pi$. 
Field intensity $1.5\times 10^{14}$ W/cm$^2$.
\label{fig4}}
\end{figure}

For the laser intensity $1\times 10^{14}$ W/cm$^2$ (the left top panel of
\Fref{fig2}), one can still discern the structures in the momentum distribution
reminiscent of the multiphoton regime. Nevertheless, the prominent global
maximum predicted by the SFA is clearly visible. This maximum takes over
completely at higher field intensities.  Each multiphoton rings visible in
\Fref{fig2} corresponds to an integer number of photons absorbed by the He atom
$
p_x^2+p_y^2=n\omega - 24.6~{\rm eV}
$
As we project the calculated 3D momentum distribution onto the $p_x,p_y$ plane,
we set $p_z=0$. The multiphoton rings are not observed in the experiment, most
probably because of the finite range of $p_z$ detected. Also, the experimental
momentum distributions \cite{Pfeiffer2012} show two symmetric lobes in the
electron momentum distribution whereas our calculations with $\phi=0$ show two
lobes of unequal strength.  This asymmetry is due to the CEP variation
investigated in \cite{Eckle2008,P.Eckle12052008} but not controlled in the later
measurements \cite{Pfeiffer2012,PhysRevLett.111.103003}.  We illustrate this
asymmetry in \Fref{fig4} where we plot the $p$-integrated momentum distributions
as functions of the angle $\theta_p$ for various CEP values.  The relative
intensity of the lobes in the second and fourth quadrants is changing with
$\phi$ in exactly the same manner as observed in
\cite{Eckle2008,P.Eckle12052008}. The figure shows some drift of the angular
maximum position $\theta_m$ with $\phi$. This is due to the drift of the
direction of the vector potential at the maximum field strength, which is
located at $t_0=3T$ when $\phi=0$ but varies slightly for other $\phi$
values. When the angular maximum values $\theta_m$ are compensated for this
drift, they are all located at the same value (9 degrees) irrespective of
$\phi$.

The offset from the SFA prediction $\theta_m=0$ can be represented in the
notations of \citet{2013arXiv1301.2766L} as
$
\theta_m=\theta_{\rm Coul}+\omega\tau \ .
$
Here $\tau$ is the tunneling time, the angle $\theta_{\rm Coul}$ arises from the
effect of the ionic potential \cite{madsen} which is neglected in the SFA.  The
TIPIS model was used in the atto-clock measurements
\cite{Pfeiffer2012,PhysRevLett.111.103003} to evaluate the Coulomb contribution
$\theta_{\rm Coul}$ and thus to evaluate the tunneling time $\tau$.

\begin{figure}[h]
%\epsfxsize=8cm
%\epsffile{PLOT.eps}\\
%\epsffile{fig5.eps}\\
 \includegraphics[width=8cm]{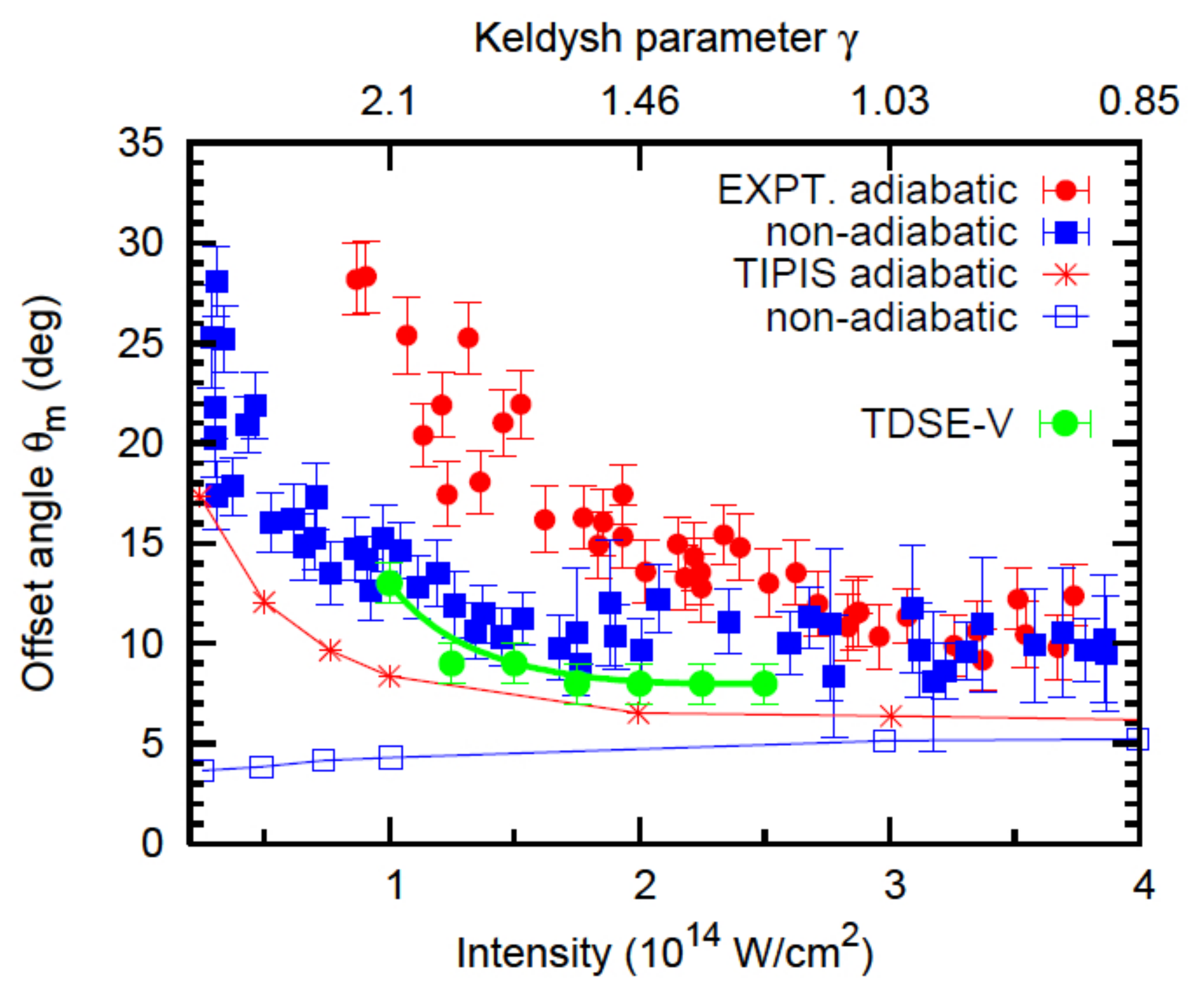}
\caption{(Color online) The offset angle $\theta_m$ of the photoelectron angular
distribution from which the tunneling time is extracted.  The two sets of the
experimental data of \citet{PhysRevLett.111.103003} corresponding to the
adiabatic and non-adiabatic field calibration are shown with the red filled
circles and blue filled squares, respectively. The two analogous sets of the
TIPIS calculations are shown with the red asterisks and blue open squares,
respectively. The present TDSE results are shown with the green filled circles.
\label{fig5}}
\end{figure}

Our numerical results for the angular offset $\theta_m$, derived from
\Fref{fig3} are shown in \Fref{fig5}.  In the same figure, we display two sets
of the experimental data of \citet{PhysRevLett.111.103003} and their
calculations using the semi-classical TIPIS model. Each set corresponds to
either adiabatic or non-adiabatic {\em in situ} calibration of the field
intensity. We see clearly that our TDSE calculations favor the set of
experimental data calibrated non-adiabatically and strongly disagree with an
alternative set of data calibrated adiabatically. At the same time, neither of
the TIPIS calculations agree with the corresponding set of the experimental
data. The adiabatic TIPIS set behaves qualitatively similar to the corresponding
set of the experimental data, but numerically is much closer to the
non-adiabatic set of the experimental. The non-adiabatic TIPIS set is
qualitatively different as it predicts the offset $\theta_m$ rising with an
increasing field intensity.

If the agreement of the present calculation with the set of experimental offset
angles, corresponding to the non-adiabatic calibration of the {\em in situ}
field intensity, is not coincidental than we can draw the  following
conclusions: (i) non-adiabatic tunneling effects are noticeable and cannot be
discarded and/or (ii) TIPIS model is inaccurate and cannot be used to extract
the tunneling time.  The second conclusion has a strong implication for the
ongoing tunneling time debate.

The authors wish to thank Claudio Cirelli and Robert Boge for their detailed and
extensive comment on the present work. Many discussions with Misha Ivanov, Olga
Smirnova and Alexandra Landsman were also very useful and stimulating.  The
authors acknowledge support of the Australian Research Council in the form of
the Discovery grant DP120101805.  Resources of the National Computational
Infrastructure (NCI) Facility were employed.

\np
%\bibliography{references,mypapers,../Where/references,../reft,../../Huillier/references,../../Huillier/hreferences}

\begin{thebibliography}{23}
\expandafter\ifx\csname natexlab\endcsname\relax\def\natexlab#1{#1}\fi
\expandafter\ifx\csname bibnamefont\endcsname\relax
  \def\bibnamefont#1{#1}\fi
\expandafter\ifx\csname bibfnamefont\endcsname\relax
  \def\bibfnamefont#1{#1}\fi
\expandafter\ifx\csname citenamefont\endcsname\relax
  \def\citenamefont#1{#1}\fi
\expandafter\ifx\csname url\endcsname\relax
  \def\url#1{\texttt{#1}}\fi
\expandafter\ifx\csname urlprefix\endcsname\relax\def\urlprefix{URL }\fi
\providecommand{\bibinfo}[2]{#2}
\providecommand{\eprint}[2][]{\url{#2}}

\bibitem[{\citenamefont{Keldysh}(1965)}]{Keldysh64}
\bibinfo{author}{\bibfnamefont{L.~V.} \bibnamefont{Keldysh}},
  \bibinfo{journal}{Sov. Phys. JETP} 
%\href{http://www.slac.stanford.edu/grp/arb/tn/arbvol5/AARD451.pdf}
{\textbf{\bibinfo{volume}{20}},
  \bibinfo{pages}{1307} (\bibinfo{year}{1965})}

\bibitem[{\citenamefont{Schultze et~al.}(2010)\citenamefont{Schultze, Fiess,
  Karpowicz, Gagnon, Korbman, Hofstetter, Neppl, Cavalieri, Komninos, Mercouris
  et~al.}}]{M.Schultze06252010}
\bibinfo{author}{\bibfnamefont{M.}~\bibnamefont{Schultze}}
%  \bibinfo{author}{\bibfnamefont{M.}~\bibnamefont{Fiess}},
%  \bibinfo{author}{\bibfnamefont{N.}~\bibnamefont{Karpowicz}},
%  \bibinfo{author}{\bibfnamefont{J.}~\bibnamefont{Gagnon}},
%  \bibinfo{author}{\bibfnamefont{M.}~\bibnamefont{Korbman}},
%  \bibinfo{author}{\bibfnamefont{M.}~\bibnamefont{Hofstetter}},
%  \bibinfo{author}{\bibfnamefont{S.}~\bibnamefont{Neppl}},
%  \bibinfo{author}{\bibfnamefont{A.~L.} \bibnamefont{Cavalieri}},
%  \bibinfo{author}{\bibfnamefont{Y.}~\bibnamefont{Komninos}},
%  \bibinfo{author}{\bibfnamefont{T.}~\bibnamefont{Mercouris}},
  \bibnamefont{et~al.}, \bibinfo{journal}{Science}
  \textbf{\bibinfo{volume}{328}}, \bibinfo{pages}{1658} (\bibinfo{year}{2010}).

\bibitem[{\citenamefont{Kl\"under et~al.}(2011)\citenamefont{Kl\"under,
  Dahlstr\"om, Gisselbrecht, Fordell, Swoboda, Gu\'enot, Johnsson, Caillat,
  Mauritsson, Maquet et~al.}}]{PhysRevLett.106.143002}
\bibinfo{author}{\bibfnamefont{K.}~\bibnamefont{Kl\"under}} %,
%  \bibinfo{author}{\bibfnamefont{J.~M.} \bibnamefont{Dahlstr\"om}},
%  \bibinfo{author}{\bibfnamefont{M.}~\bibnamefont{Gisselbrecht}},
%  \bibinfo{author}{\bibfnamefont{T.}~\bibnamefont{Fordell}},
%  \bibinfo{author}{\bibfnamefont{M.}~\bibnamefont{Swoboda}},
%  \bibinfo{author}{\bibfnamefont{D.}~\bibnamefont{Gu\'enot}},
%  \bibinfo{author}{\bibfnamefont{P.}~\bibnamefont{Johnsson}},
%  \bibinfo{author}{\bibfnamefont{J.}~\bibnamefont{Caillat}},
%  \bibinfo{author}{\bibfnamefont{J.}~\bibnamefont{Mauritsson}},
%  \bibinfo{author}{\bibfnamefont{A.}~\bibnamefont{Maquet}},
  \bibnamefont{et~al.}, \bibinfo{journal}{Phys. Rev. Lett.}
%\href{http://dx.doi.org/10.1103/PhysRevLett.106.143002}
{  \textbf{\bibinfo{volume}{106}}, \bibinfo{pages}{143002}
  (\bibinfo{year}{2011})}.

\bibitem[{\citenamefont{Gu\'enot et~al.}(2012)\citenamefont{Gu\'enot,
  Kl\"under, Arnold, Kroon, Dahlstr\"om, Miranda, Fordell, Gisselbrecht,
  Johnsson, Mauritsson et~al.}}]{PhysRevA.85.053424}
\bibinfo{author}{\bibfnamefont{D.}~\bibnamefont{Gu\'enot}} %,
%  \bibinfo{author}{\bibfnamefont{K.}~\bibnamefont{Kl\"under}},
%  \bibinfo{author}{\bibfnamefont{C.~L.} \bibnamefont{Arnold}},
%  \bibinfo{author}{\bibfnamefont{D.}~\bibnamefont{Kroon}},
%  \bibinfo{author}{\bibfnamefont{J.~M.} \bibnamefont{Dahlstr\"om}},
%  \bibinfo{author}{\bibfnamefont{M.}~\bibnamefont{Miranda}},
%  \bibinfo{author}{\bibfnamefont{T.}~\bibnamefont{Fordell}},
%  \bibinfo{author}{\bibfnamefont{M.}~\bibnamefont{Gisselbrecht}},
%  \bibinfo{author}{\bibfnamefont{P.}~\bibnamefont{Johnsson}},
%  \bibinfo{author}{\bibfnamefont{J.}~\bibnamefont{Mauritsson}},
  \bibnamefont{et~al.}, 
%  \emph{\bibinfo{title}{Photoemission-time-delay
%  measurements and calculations close to the 3$s$-ionization-cross-section
%  minimum in {Ar}}}, 
\bibinfo{journal}{Phys. Rev. A}
  %\href{http://link.aps.org/doi/10.1103/PhysRevA.85.053424}
{
  \textbf{\bibinfo{volume}{85}}, \bibinfo{pages}{053424}
  (\bibinfo{year}{2012})}.



\bibitem[{\citenamefont{de~Carvalho and Nussenzveig}(2002)}]{deCarvalho200283}
\bibinfo{author}{\bibfnamefont{C.~A.~A.} \bibnamefont{de~Carvalho}}
  \bibnamefont{and} \bibinfo{author}{\bibfnamefont{H.~M.}
  \bibnamefont{Nussenzveig}}, \bibinfo{journal}{Phys. Rep.}
 %\href{http://www.sciencedirect.com/science/article/B6TVP-44NM47M-1/%
%2/8956c9ddffa3de8f24ca93f1dc65a47e}
{  \textbf{\bibinfo{volume}{364}}, \bibinfo{pages}{83 } (\bibinfo{year}{2002})}.

\bibitem[{\citenamefont{Kheifets}(2013)}]{PhysRevA.87.063404}
\bibinfo{author}{\bibfnamefont{A.~S.} \bibnamefont{Kheifets}},
  \bibinfo{journal}{Phys. Rev. A} 
  %\href{http://link.aps.org/doi/10.1103/PhysRevA.87.063404}
{\textbf{\bibinfo{volume}{87}},
  \bibinfo{pages}{063404} (\bibinfo{year}{2013})}

\bibitem[{\citenamefont{Eckle et~al.}(2008{\natexlab{a}})\citenamefont{Eckle,
  Smolarski, Schlup, Biegert, Staudte, Schoffler, Muller, Dorner, and
  Keller}}]{Eckle2008}
\bibinfo{author}{\bibfnamefont{P.}~\bibnamefont{Eckle}} %,
%  \bibinfo{author}{\bibfnamefont{M.}~\bibnamefont{Smolarski}},
%  \bibinfo{author}{\bibfnamefont{P.}~\bibnamefont{Schlup}},
%  \bibinfo{author}{\bibfnamefont{J.}~\bibnamefont{Biegert}},
%  \bibinfo{author}{\bibfnamefont{A.}~\bibnamefont{Staudte}},
%  \bibinfo{author}{\bibfnamefont{M.}~\bibnamefont{Schoffler}},
%  \bibinfo{author}{\bibfnamefont{H.~G.} \bibnamefont{Muller}},
%  \bibinfo{author}{\bibfnamefont{R.}~\bibnamefont{Dorner}}, \bibnamefont{and}
%  \bibinfo{author}{\bibfnamefont{U.}~\bibnamefont{Keller}},
 \bibnamefont{et~al.},  
  \bibinfo{journal}{Nature Phys.} 
%\href{http://dx.doi.org/10.1038/nphys982}
{\textbf{\bibinfo{volume}{4}},
  \bibinfo{pages}{565} (\bibinfo{year}{2008}{\natexlab{a}})}.


\bibitem[{\citenamefont{Eckle et~al.}(2008)\citenamefont{Eckle, Pfeiffer,
  Cirelli, Staudte, Dorner, Muller, Buttiker, and Keller}}]{P.Eckle12052008}
\bibinfo{author}{\bibfnamefont{P.}~\bibnamefont{Eckle}} %,
%  \bibinfo{author}{\bibfnamefont{A.~N.} \bibnamefont{Pfeiffer}},
%  \bibinfo{author}{\bibfnamefont{C.}~\bibnamefont{Cirelli}},
%  \bibinfo{author}{\bibfnamefont{A.}~\bibnamefont{Staudte}},
%  \bibinfo{author}{\bibfnamefont{R.}~\bibnamefont{Dorner}},
%  \bibinfo{author}{\bibfnamefont{H.~G.} \bibnamefont{Muller}},
%  \bibinfo{author}{\bibfnamefont{M.}~\bibnamefont{Buttiker}}, \bibnamefont{and}
%  \bibinfo{author}{\bibfnamefont{U.}~\bibnamefont{Keller}},
 \bibnamefont{et~al.},  
\bibinfo{journal}{Science} 
%\href{http://dx.doi.org/10.1126/science.1163439}
{\textbf{\bibinfo{volume}{322}},
  \bibinfo{pages}{1525} (\bibinfo{year}{2008})}.

\bibitem[{\citenamefont{Shafir et~al.}(2012)\citenamefont{Shafir, Soifer,
  Bruner, Dagan, Mairesse, Patchkovskii, Ivanov, Smirnova, and
  Dudovich}}]{Shafir2012}
\bibinfo{author}{\bibfnamefont{D.}~\bibnamefont{Shafir}} %,
%  \bibinfo{author}{\bibfnamefont{H.}~\bibnamefont{Soifer}},
%  \bibinfo{author}{\bibfnamefont{B.~D.} \bibnamefont{Bruner}},
%  \bibinfo{author}{\bibfnamefont{M.}~\bibnamefont{Dagan}},
%  \bibinfo{author}{\bibfnamefont{Y.}~\bibnamefont{Mairesse}},
%  \bibinfo{author}{\bibfnamefont{S.}~\bibnamefont{Patchkovskii}},
%  \bibinfo{author}{\bibfnamefont{M.~Y.} \bibnamefont{Ivanov}},
%  \bibinfo{author}{\bibfnamefont{O.}~\bibnamefont{Smirnova}}, \bibnamefont{and}
%  \bibinfo{author}{\bibfnamefont{N.}~\bibnamefont{Dudovich}},
 \bibnamefont{et~al.},  
  \bibinfo{journal}{Nature} 
  %\href{http://dx.doi.org/10.1038/nature11025}
{\textbf{\bibinfo{volume}{485}},
  \bibinfo{pages}{343} (\bibinfo{year}{2012})}.

\bibitem[{\citenamefont{Lein}(2012)}]{Lein2012}
\bibinfo{author}{\bibfnamefont{M.}~\bibnamefont{Lein}},
  \bibinfo{journal}{Nature} 
  %\href{http://dx.doi.org/10.1038/485313a}
{\textbf{\bibinfo{volume}{485}},
  \bibinfo{pages}{313} (\bibinfo{year}{2012})}.


\bibitem[{\citenamefont{Landauer and Martin}(1994)}]{RevModPhys.66.217}
\bibinfo{author}{\bibfnamefont{R.}~\bibnamefont{Landauer}} \bibnamefont{and}
  \bibinfo{author}{\bibfnamefont{T.}~\bibnamefont{Martin}},
  \bibinfo{journal}{Rev. Mod. Phys.} 
  %\href{http://link.aps.org/doi/10.1103/RevModPhys.66.217}
{\textbf{\bibinfo{volume}{66}},
  \bibinfo{pages}{217} (\bibinfo{year}{1994})}.

\bibitem[{\citenamefont{Pfeiffer
  et~al.}(2012{\natexlab{a}})\citenamefont{Pfeiffer, Cirelli, Smolarski,
  Dimitrovski, Abu-samha, Madsen, and Keller}}]{Pfeiffer2012}
\bibinfo{author}{\bibfnamefont{A.~N.}   \bibnamefont{Pfeiffer}} %,
%  \bibinfo{author}{\bibfnamefont{C.}~\bibnamefont{Cirelli}},
%  \bibinfo{author}{\bibfnamefont{M.}~\bibnamefont{Smolarski}},
%  \bibinfo{author}{\bibfnamefont{D.}~\bibnamefont{Dimitrovski}},
%  \bibinfo{author}{\bibfnamefont{M.}~\bibnamefont{Abu-samha}},
%  \bibinfo{author}{\bibfnamefont{L.~B.} \bibnamefont{Madsen}},
%  \bibnamefont{and} \bibinfo{author}{\bibfnamefont{U.}~\bibnamefont{Keller}},
 \bibnamefont{et~al.},  
  \bibinfo{journal}{Nature Phys.} 
  %\href{http://dx.doi.org/10.1038/nphys2125}
{\textbf{\bibinfo{volume}{8}}, \bibinfo{pages}{76}
  (\bibinfo{year}{2012}{\natexlab{a}})}.

\bibitem[{\citenamefont{Pfeiffer
  et~al.}(2012{\natexlab{b}})\citenamefont{Pfeiffer, Cirelli, Landsman,
  Smolarski, Dimitrovski, Madsen, and Keller}}]{PhysRevLett.109.083002}
\bibinfo{author}{\bibfnamefont{A.~N.}   \bibnamefont{Pfeiffer}} %,
%  \bibinfo{author}{\bibfnamefont{C.}~\bibnamefont{Cirelli}},
%  \bibinfo{author}{\bibfnamefont{A.~S.} \bibnamefont{Landsman}},
%  \bibinfo{author}{\bibfnamefont{M.}~\bibnamefont{Smolarski}},
%  \bibinfo{author}{\bibfnamefont{D.}~\bibnamefont{Dimitrovski}},
%  \bibinfo{author}{\bibfnamefont{L.~B.} \bibnamefont{Madsen}},
%  \bibnamefont{and} \bibinfo{author}{\bibfnamefont{U.}~\bibnamefont{Keller}},
 \bibnamefont{et~al.},  
  \bibinfo{journal}{Phys. Rev. Lett.}
  %\href{http://link.aps.org/doi/10.1103/PhysRevLett.109.083002}
{ \textbf{\bibinfo{volume}{109}},
  \bibinfo{pages}{083002} (\bibinfo{year}{2012}{\natexlab{b}})}.

\bibitem[{\citenamefont{Hofmann et~al.}(2013)\citenamefont{Hofmann, Landsman,
  Cirelli, Pfeiffer, and Keller}}]{0953-4075-46-12-125601}
\bibinfo{author}{\bibfnamefont{C.}~\bibnamefont{Hofmann}},
  \bibinfo{author}{\bibfnamefont{A.~S.} \bibnamefont{Landsman}},
  \bibinfo{author}{\bibfnamefont{C.}~\bibnamefont{Cirelli}},
  \bibinfo{author}{\bibfnamefont{A.~N.} \bibnamefont{Pfeiffer}},
  \bibnamefont{and} \bibinfo{author}{\bibfnamefont{U.}~\bibnamefont{Keller}},
  \bibinfo{journal}{J.~Phys.~B} 
  %\href{http://stacks.iop.org/0953-4075/46/i=12/a=125601}
{\textbf{\bibinfo{volume}{46}},
  \bibinfo{pages}{125601} (\bibinfo{year}{2013})}.

\bibitem[{\citenamefont{{Weger} et~al.}(2013)\citenamefont{{Weger}, {Maurer},
  {Ludwig}, {Gallmann}, and {Keller}}}]{2013arXiv1306.6280W}
\bibinfo{author}{\bibfnamefont{M.}~\bibnamefont{{Weger}}} %,
%  \bibinfo{author}{\bibfnamefont{J.}~\bibnamefont{{Maurer}}},
%  \bibinfo{author}{\bibfnamefont{A.}~\bibnamefont{{Ludwig}}},
%  \bibinfo{author}{\bibfnamefont{L.}~\bibnamefont{{Gallmann}}},
%  \bibnamefont{and} \bibinfo{author}{\bibfnamefont{U.}~\bibnamefont{{Keller}}},
 \bibnamefont{et~al.},  
  \bibinfo{journal}{ArXiv e-prints} 
%\href{http://adsabs.harvard.edu/abs/2013arXiv1301.2766L}
{   \eprint{1306.6280} (\bibinfo{year}{2013})}.

\bibitem[{\citenamefont{{Landsman} et~al.}(2013)\citenamefont{{Landsman},
  {Weger}, {Maurer}, {Boge}, {Ludwig}, {Heuser}, {Cirelli}, {Gallmann}, and
  {Keller}}}]{2013arXiv1301.2766L}
\bibinfo{author}{\bibfnamefont{A.}~\bibnamefont{{Landsman}}}  %,
%  \bibinfo{author}{\bibfnamefont{M.}~\bibnamefont{{Weger}}},
%  \bibinfo{author}{\bibfnamefont{J.}~\bibnamefont{{Maurer}}},
%  \bibinfo{author}{\bibfnamefont{R.}~\bibnamefont{{Boge}}},
%  \bibinfo{author}{\bibfnamefont{A.}~\bibnamefont{{Ludwig}}},
%  \bibinfo{author}{\bibfnamefont{S.}~\bibnamefont{{Heuser}}},
%  \bibinfo{author}{\bibfnamefont{C.}~\bibnamefont{{Cirelli}}},
%  \bibinfo{author}{\bibfnamefont{L.}~\bibnamefont{{Gallmann}}},
%  \bibnamefont{and} \bibinfo{author}{\bibfnamefont{U.}~\bibnamefont{{Keller}}},
 \bibnamefont{et~al.},  
  \bibinfo{journal}{ArXiv e-prints}
%\href{http://adsabs.harvard.edu/abs/2013arXiv1306.6280W}
{    \eprint{1301.2766} (\bibinfo{year}{2013})}.

\bibitem[{\citenamefont{Boge et~al.}(2013)\citenamefont{Boge, Cirelli,
  Landsman, Heuser, Ludwig, Maurer, Weger, Gallmann, and
  Keller}}]{PhysRevLett.111.103003}
\bibinfo{author}{\bibfnamefont{R.}~\bibnamefont{Boge}} %,
%  \bibinfo{author}{\bibfnamefont{C.}~\bibnamefont{Cirelli}},
%  \bibinfo{author}{\bibfnamefont{A.~S.} \bibnamefont{Landsman}},
%  \bibinfo{author}{\bibfnamefont{S.}~\bibnamefont{Heuser}},
%  \bibinfo{author}{\bibfnamefont{A.}~\bibnamefont{Ludwig}},
%  \bibinfo{author}{\bibfnamefont{J.}~\bibnamefont{Maurer}},
%  \bibinfo{author}{\bibfnamefont{M.}~\bibnamefont{Weger}},
%  \bibinfo{author}{\bibfnamefont{L.}~\bibnamefont{Gallmann}}, \bibnamefont{and}
%  \bibinfo{author}{\bibfnamefont{U.}~\bibnamefont{Keller}},
 \bibnamefont{et~al.},  
  \bibinfo{journal}{Phys. Rev. Lett.} 
  %\href{http://link.aps.org/doi/10.1103/PhysRevLett.111.103003}.
{\textbf{\bibinfo{volume}{111}},
  \bibinfo{pages}{103003} (\bibinfo{year}{2013})}.

\bibitem[{\citenamefont{{Cirelli, C. and Boge, R.}}(2013)}]{Cirelli2013}
\bibinfo{author}{\bibnamefont{{C. Cirelli and R. Boge}}}
  (\bibinfo{year}{2013}), \bibinfo{note}{private communication}.

\bibitem[{\citenamefont{Yudin and Ivanov}(2001)}]{PhysRevA.64.013409}
\bibinfo{author}{\bibfnamefont{G.~L.} \bibnamefont{Yudin}} \bibnamefont{and}
  \bibinfo{author}{\bibfnamefont{M.~Y.} \bibnamefont{Ivanov}},
  \bibinfo{journal}{Phys. Rev. A} 
  %\href{http://link.aps.org/doi/10.1103/PhysRevA.64.013409}
{\textbf{\bibinfo{volume}{64}},
  \bibinfo{pages}{013409} (\bibinfo{year}{2001})}.


%% \bibitem[{\citenamefont{B\"uttiker}(1983)}]{PhysRevB.27.6178}
%% \bibinfo{author}{\bibfnamefont{M.}~\bibnamefont{B\"uttiker}},
%%   \bibinfo{journal}{Phys. Rev. B} 
%%   %\href{http://link.aps.org/doi/10.1103/PhysRevB.27.6178}
%% {\textbf{\bibinfo{volume}{27}},
%%   \bibinfo{pages}{6178} (\bibinfo{year}{1983})}.

%% \bibitem[{\citenamefont{Fertig}(1990)}]{PhysRevLett.65.2321}
%% \bibinfo{author}{\bibfnamefont{H.~A.} \bibnamefont{Fertig}},
%%   \bibinfo{journal}{Phys. Rev. Lett.} 
%%   %\href{http://link.aps.org/doi/10.1103/PhysRevLett.65.2321}
%% {\textbf{\bibinfo{volume}{65}},
%%   \bibinfo{pages}{2321} (\bibinfo{year}{1990})}.

%% \bibitem[{\citenamefont{Yamada}(2004)}]{PhysRevLett.93.170401}
%% \bibinfo{author}{\bibfnamefont{N.}~\bibnamefont{Yamada}},
%%   \bibinfo{journal}{Phys. Rev. Lett.} 
%%   %\href{http://link.aps.org/doi/10.1103/PhysRevLett.93.170401}
%% {\textbf{\bibinfo{volume}{93}},
%%   \bibinfo{pages}{170401} (\bibinfo{year}{2004})}.


\bibitem[{\citenamefont{Sarsa et~al.}(2004)\citenamefont{Sarsa, G\'{a}lvez, and
  Buendia}}]{Sarsa2004163}
\bibinfo{author}{\bibfnamefont{A.}~\bibnamefont{Sarsa}},
  \bibinfo{author}{\bibfnamefont{F.~J.} \bibnamefont{G\'{a}lvez}},
  \bibnamefont{and} \bibinfo{author}{\bibfnamefont{E.}~\bibnamefont{Buendia}},
  \bibinfo{journal}{Atomic Data and Nuclear Data Tables}
%\href{http://www.sciencedirect.com/science/article/B6WBB-4D7CCCM-1/2/4aac5695d2170326a9834bb19540d0b1}  
{  \textbf{\bibinfo{volume}{88}}, \bibinfo{pages}{163 } (\bibinfo{year}{2004})}.


\bibitem[{\citenamefont{Green et~al.}(1969)\citenamefont{Green, Sellin, and
  Zachor}}]{PhysRev.184.1}
\bibinfo{author}{\bibfnamefont{A.~E.~S.} \bibnamefont{Green}},
  \bibinfo{author}{\bibfnamefont{D.~L.} \bibnamefont{Sellin}},
  \bibnamefont{and} \bibinfo{author}{\bibfnamefont{A.~S.}
  \bibnamefont{Zachor}}, \bibinfo{journal}{Phys. Rev.}
  %\href{http://link.aps.org/doi/10.1103/PhysRev.184.1}
{  \textbf{\bibinfo{volume}{184}}, \bibinfo{pages}{1} (\bibinfo{year}{1969})}.



\bibitem[{\citenamefont{Nurhuda and
  Faisal}(1999{\natexlab{a}})}]{PhysRevA.60.3125}
\bibinfo{author}{\bibfnamefont{M.}~\bibnamefont{Nurhuda}} \bibnamefont{and}
  \bibinfo{author}{\bibfnamefont{F.~H.~M.} \bibnamefont{Faisal}},
  \bibinfo{journal}{Phys.~Rev.~A} 
  %\href{http://dx.doi.org/10.1103/PhysRevA.60.3125}
{\textbf{\bibinfo{volume}{60}},
  \bibinfo{pages}{3125} (\bibinfo{year}{1999}{\natexlab{a}})}.

\bibitem[{\citenamefont{{Grum-Grzhimailo {\em et al}}}(2010)}]{klaus1}
\bibinfo{author}{\bibfnamefont{A.~N.} \bibnamefont{{Grum-Grzhimailo {\em et
  al}}}}, \bibinfo{journal}{Phys. Rev. A}
  %\href{http://dx.doi.org/10.1103/PhysRevA.81.043408}
{ \textbf{\bibinfo{volume}{81}},
  \bibinfo{pages}{043408} (\bibinfo{year}{2010})}.

\bibitem[{\citenamefont{Ivanov}(2011)}]{dstrong}
\bibinfo{author}{\bibfnamefont{I.~A.} \bibnamefont{Ivanov}},
  \bibinfo{journal}{Phys. Rev. A} 
  %\href{http://dx.doi.org/10.1103/PhysRevA.83.023421}
{\textbf{\bibinfo{volume}{83}},
  \bibinfo{pages}{023421} (\bibinfo{year}{2011})}.

\bibitem[{\citenamefont{Ivanov and Kheifets}(2013)}]{PhysRevA.87.033407}
\bibinfo{author}{\bibfnamefont{I.~A.} \bibnamefont{Ivanov}} \bibnamefont{and}
  \bibinfo{author}{\bibfnamefont{A.~S.} \bibnamefont{Kheifets}},
  \bibinfo{journal}{Phys. Rev. A} 
  %\href{http://link.aps.org/doi/10.1103/PhysRevA.87.033407}
{\textbf{\bibinfo{volume}{87}},
  \bibinfo{pages}{033407} (\bibinfo{year}{2013})}.

\bibitem[{\citenamefont{Nurhuda and Faisal}(1999{\natexlab{b}})}]{velocity1}
\bibinfo{author}{\bibfnamefont{M.}~\bibnamefont{Nurhuda}} \bibnamefont{and}
  \bibinfo{author}{\bibfnamefont{F.~H.~M.} \bibnamefont{Faisal}},
  \bibinfo{journal}{Phys. Rev. A}
%\href{http://dx.doi.org/10.1103/PhysRevA.60.3125}
{ \textbf{\bibinfo{volume}{60}},
  \bibinfo{pages}{3125} (\bibinfo{year}{1999}{\natexlab{b}})}.

\bibitem[{\citenamefont{Perelomov and Popov}(1967)}]{Perelomov67}
\bibinfo{author}{\bibfnamefont{A.~M.} \bibnamefont{Perelomov}}
  \bibnamefont{and} \bibinfo{author}{\bibfnamefont{V.~S.} \bibnamefont{Popov}},
  \bibinfo{journal}{Sov. Phys. JETP} 
%\href{http://www.jetp.ac.ru/cgi-bin/e/index/e/25/2/p336?a=list}
{\textbf{\bibinfo{volume}{25}},
  \bibinfo{pages}{336} (\bibinfo{year}{1967})}.


\bibitem[{\citenamefont{Madsen et~al.}(2007)\citenamefont{Madsen, Nikolopoulos,
  Kjeldsen, and Fern\'andez}}]{madsen}
\bibinfo{author}{\bibfnamefont{L.~B.} \bibnamefont{Madsen}},
  \bibinfo{author}{\bibfnamefont{L.~A.~A.} \bibnamefont{Nikolopoulos}},
  \bibinfo{author}{\bibfnamefont{T.~K.} \bibnamefont{Kjeldsen}},
  \bibnamefont{and}
  \bibinfo{author}{\bibfnamefont{J.}~\bibnamefont{Fern\'andez}},
  \bibinfo{journal}{Phys. Rev. A} 
%\href{http://pra.aps.org/abstract/PRA/v76/i6/e063407}
{\textbf{\bibinfo{volume}{76}},
  \bibinfo{pages}{063407} (\bibinfo{year}{2007})}.

\end{thebibliography}
%\end{document}

\end{document}